\documentclass[%
 reprint,
superscriptaddress,
 amsmath,amssymb,
 aps,
]{revtex4-2}
\usepackage{color}
\usepackage{gensymb}
\usepackage{graphicx}
\usepackage{dcolumn}
\usepackage{bm}
\usepackage{hyperref}
\usepackage{epstopdf}
\usepackage{float}
\usepackage{listings}
\usepackage{times,mathptm,bm}
\usepackage{xcolor}
\usepackage[export]{adjustbox}
\usepackage{upgreek}

\input ulem.sty\relax
\catcode`\@=11 %

\font\uwavefont=lasyb10 scaled 652

\def\uwave{%
  \bgroup
    \markoverwith{%
      \lower3.5\p@\hbox{\uwavefont\char58}%
    }%
  \ULon
}

\begin{document}

\title{Early Stages of Drop Coalescence}

\author{Antoine Deblais}
\affiliation{Van der Waals-Zeeman Institute, Institute of Physics, University of Amsterdam, 1098XH Amsterdam, The Netherlands.}
\author{Kaili Xie}
\affiliation{Van der Waals-Zeeman Institute, Institute of Physics, University of Amsterdam, 1098XH Amsterdam, The Netherlands.}
\author{Peter~Lewin-Jones}
\affiliation{Mathematics Institute, University of Warwick, Coventry CV4 7AL, United Kingdom.}
\author{Dirk~Aarts}
\affiliation{Department of Chemistry, Physical and Theoretical Chemistry Laboratory, University of Oxford,
South Parks Road, Oxford OX1 3QZ, United Kingdom.}
\author{Miguel~A.~Herrada }
\affiliation{Escuela Tecnica Superior de Ingeneria, Universidad de Sevilla, Seville 41092, Spain}
\author{Jens~Eggers}
\affiliation{School of Mathematics, University of Bristol, University Walk, Bristol BS8 1 TW, United Kingdom.}
\author{James~E.~Sprittles}
\affiliation{Mathematics Institute, University of Warwick, Coventry CV4 7AL, United Kingdom.}
\author{Daniel~Bonn}
\affiliation{Van der Waals-Zeeman Institute, Institute of Physics, University of Amsterdam, 1098XH Amsterdam, The Netherlands.}
\date{\today}

\begin{abstract}
Despite the large body of research on coalescence, firm agreement between experiment, theory, and computation has not been established for the very first moments following the initial contact of two liquid volumes. Combining a range of experimental and computational modeling approaches in two different geometries, namely drop-drop and drop-bath configurations, we have been able to elucidate the influence of the intervening gas and van der Waals forces on coalescence. For simple liquids considered here, the gas influences both pre- and post-contact regimes, with jump-to-contact being the primary mode of merging. Subsequently, wave-like air pockets are observed and ultimately influence the initial opening dynamics of the neck.
\end{abstract}

\pacs{Valid PACS appear here}
\maketitle

The merging of liquid volumes is important for a plethora of fluid phenomena: for example, drop-drop coalescence dynamics dictate the size of raindrops, efficacy of virus transmission, accuracy of inkjet printing and efficiency of spraying phenomena \cite{lohse2022,glasser2019,liu2014,somsen2020,kooij2022}, whilst drop-surface interactions are key for production of ocean mist, air-sea gas exchange and airborne salt particles \cite{wanninkhof2009,raes2000}.

When colliding drops are able to displace the surrounding gas to touch, coalescence is initiated by the formation of a liquid bridge (or neck) between the two drops, which subsequently grows with surface tension driving the expansion. Depending on the parameter regime, namely whether it is the viscosity or inertia of the drop that resists the capillary forces, different scaling have been derived for this growth \cite{eggers1999,aarts2005}. 

In the conventional hydrodynamic description of coalescence, growth is initially inhibited by the viscosity $\eta$ of the liquid. The rate of growth is then determined by a balance of capillary and viscous forces, and a simple dimensional analysis then leads directly to a capillary velocity $U_{cap} \approx \gamma/\eta$, with $\gamma$ the surface tension, which gives the rate at which the fluid bridge increases. A more detailed analysis \cite{eggers1999} shows that there is a logarithmic correction to this, leading to a temporal variation of the bridge radius $R \propto \tau\log \tau$, where $\tau$ = $t$ - $t_0$ is the time from initial contact $t_0$. 

For the coalescence of water droplets ($\gamma \approx$ 70 mN/m and $\eta$ = 1 mPa$\cdot$s), $U_{cap} \approx$~70 m/s. This means that very rapidly the fluid inertia becomes important; the balance between inertial and viscous forces is dictated by the local Reynolds number $Re(\tau)= \rho \gamma R^2 / R_0 \eta^2$ with $\rho$ the density, $R(\tau)$ the radius of the neck and $R_0$ the radius of the drop \cite{paulsen2011}. For millimeter-sized drops of water, $Re \approx 1$ as soon as $R \approx 1 \mu m$. Thus, for water, the viscous regime is hard to observe, and the coalescence is, from a practical point of view, completely inertial. The inertial dynamics is then given by 
\begin{equation}
R(\tau) = A {\left(\frac{\gamma R_0}{\rho}\right)}^{1/4} \tau^{1/2},
\end{equation}
where the prefactor $A$ covers 1-1.25 from imaging experiment measurement~\cite{wu2004,aarts2005,thoroddsen2005}, slightly lower than 1.62 from numerical simulation \cite{duchemin2003}. For coalescence, simulations and classical theories assume that the initial state consists of two stationary undeformed spherical drops joining with a very small liquid bridge, with the Laplace pressure for a vanishing contact radius going to infinity. This poses the question how these drops actually  come into contact and displace the intervening gas?  

Experimentally, this is a very difficult question as direct imaging is evidently limited by both the optical resolution and the rate at which images can be acquired \cite{wu2004,aarts2005}. In addition, focusing on the plane of the coalescing drops and `seeing into' its cusp-like region may lead to spurious exponents for the dynamics, as shown in \cite{eddi2013}. To overcome these issues, Paulsen \textit{et al.} \cite{paulsen2011,paulsen2013} deployed an electrical method to infer the radius of the bridge in the early stages and proposed a new initial regime in which inertia, capillarity, and viscosity are all important. However, simulations of the full Navier-Stokes system \cite{sprittles2014,anthony2020} show no evidence for this regime, so the links between experiment and theory remain unclear.

In this Letter, we study the initial stages of inertial drop coalescence at extremely short times ($\approx 0.1~\mu$s) and small length scales ($\approx 2~\mu$m) in two configurations, drop-drop and drop-bath, elucidating the role of initial contact and air drainage in the dynamics of neck opening. In the drop-drop configuration, we show that with good control of initial conditions in the electrical setup, the dynamics of the neck opening follows the expected inertial scaling without deviation, suggesting that previous experimental discrepancies were due to initial interface deformation caused by undesired charges. In the drop-bath configuration, we developed a new imaging setup that allows us to directly visualize the air-liquid interface evolution over time. We show for the first time that the rupture of an air film during the merging of a drop with a bulk liquid occurs with a jump-to-contact, generating capillary waves at the liquid surface ahead of an entrapped air pocket. This consequently affects the inertial opening dynamics at the very first moments after coalescence. In both cases, our experimental observations agree well with direct simulations that incorporate both liquid and gas phases without assuming an initially spherical shape; instead, we pre-compute the initial conditions based on the approach problem, allowing us to accurately model the two different configurations.

\begin{figure}[t]
    \centering
    \includegraphics[width=1\columnwidth]{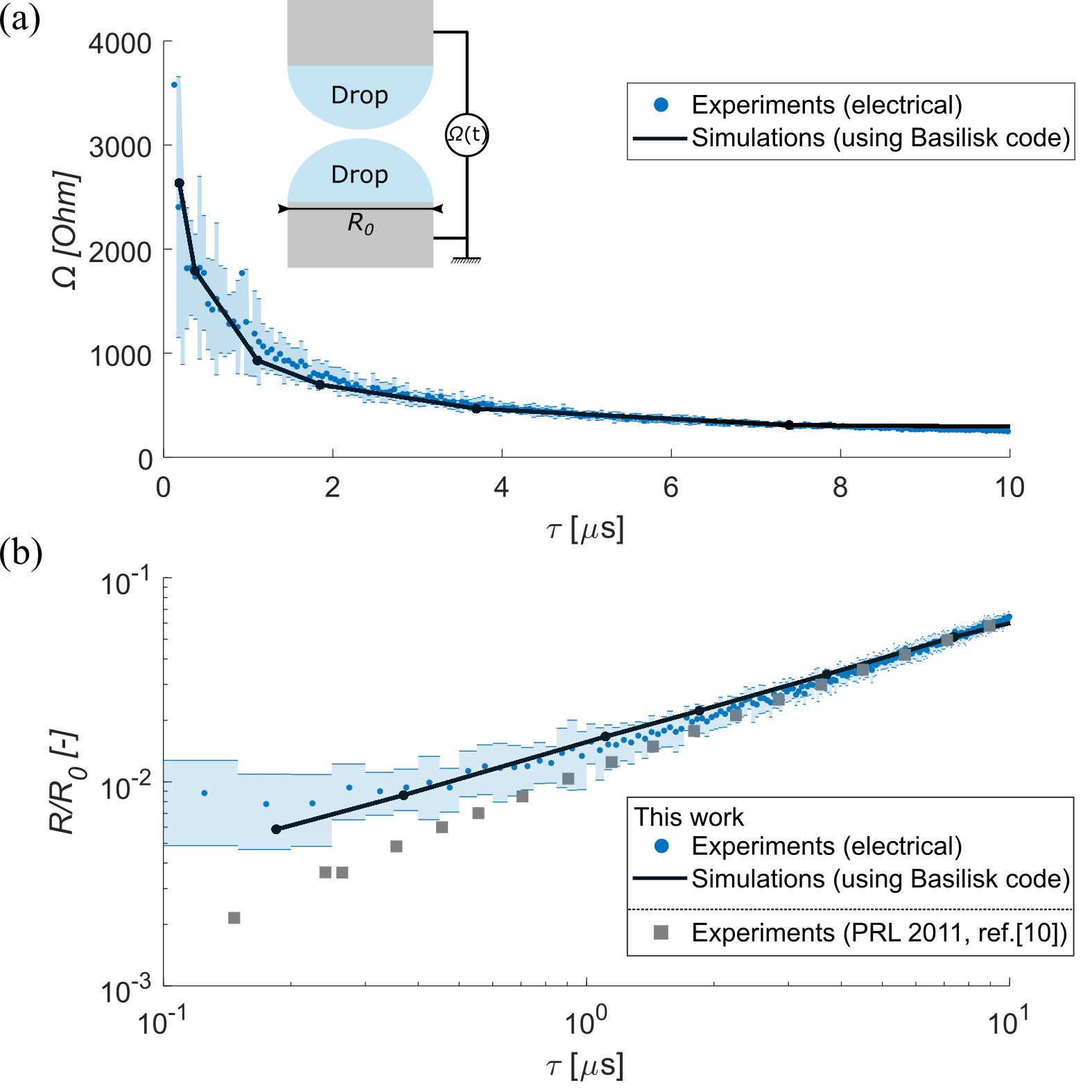}
    \caption{(a)~Electrical resistance $\Omega$ of a low-viscous saline capillary bridge ($\eta$ = 1.8 mPa s) between two electrodes ($R_0$ = 1 mm) as a function of time. The upper electrode moves at a very low constant downward speed, initiating coalescence; the electrical resistance is obtained from the complex impedance of the circuit, following the method detailed in \cite{paulsen2013}. (b)~From the electrical resistance shown in (a), the opening dynamics of the neck can be calculated by Eq.~\ref{eq:radius_electric}. Data from Ref.~\cite{paulsen2011} are also plotted for comparison (viscosity $\eta$ = 1.9 mPa$\cdot$s). Dots represent averaged values for 3 different experiments, while the shaded area corresponds to standard deviations.}
    \label{fig:electric}
\end{figure}

\textit{Drop-Drop.} We use an electrical technique similar to those described in \cite{case2009,paulsen2011,paulsen2012,paulsen2013} to measure the coalescence between two drops. The experiments consist of two electrodes facing each other, each holding a liquid droplet. The upper electrode is translated downward using a speed-controlled translation stage. To increase the spatial and temporal resolution of this measurement, we use salt water (NaCl) with high conductivity (30 mS/cm, see Sup.~Table~1). This allows us to achieve micron spatial and nearly nanosecond temporal resolution, and also avoids the need for large applied electrical tension between the electrodes, ensuring that the shape of the droplet remains spherical rather than conical since the initial shape of the droplet is known to strongly influence the dynamics of the neck opening \cite{bird2009,eddi2013}.

In Fig.~\ref{fig:electric}(a) we plot the measured conductivity as a function of time. The shape of the liquid bridge is computed from a simulation, allowing us to calculate the radius of the neck over time. We used a numerical scheme based on a volume-of-fluid (VOF) method provided by the `Basilisk' software \cite{basilisk} which uses QUADTREES \cite{popinet2011} to allow efficient adaptive grid refinement of the interface close to the coalescence region (Sup. Mat.~\cite{Sup}). The measured resistance of the liquid bridge data is well described by the simulations at early stage of the coalescence point.

\begin{figure}[b]
    \centering
    \includegraphics[width=\columnwidth]{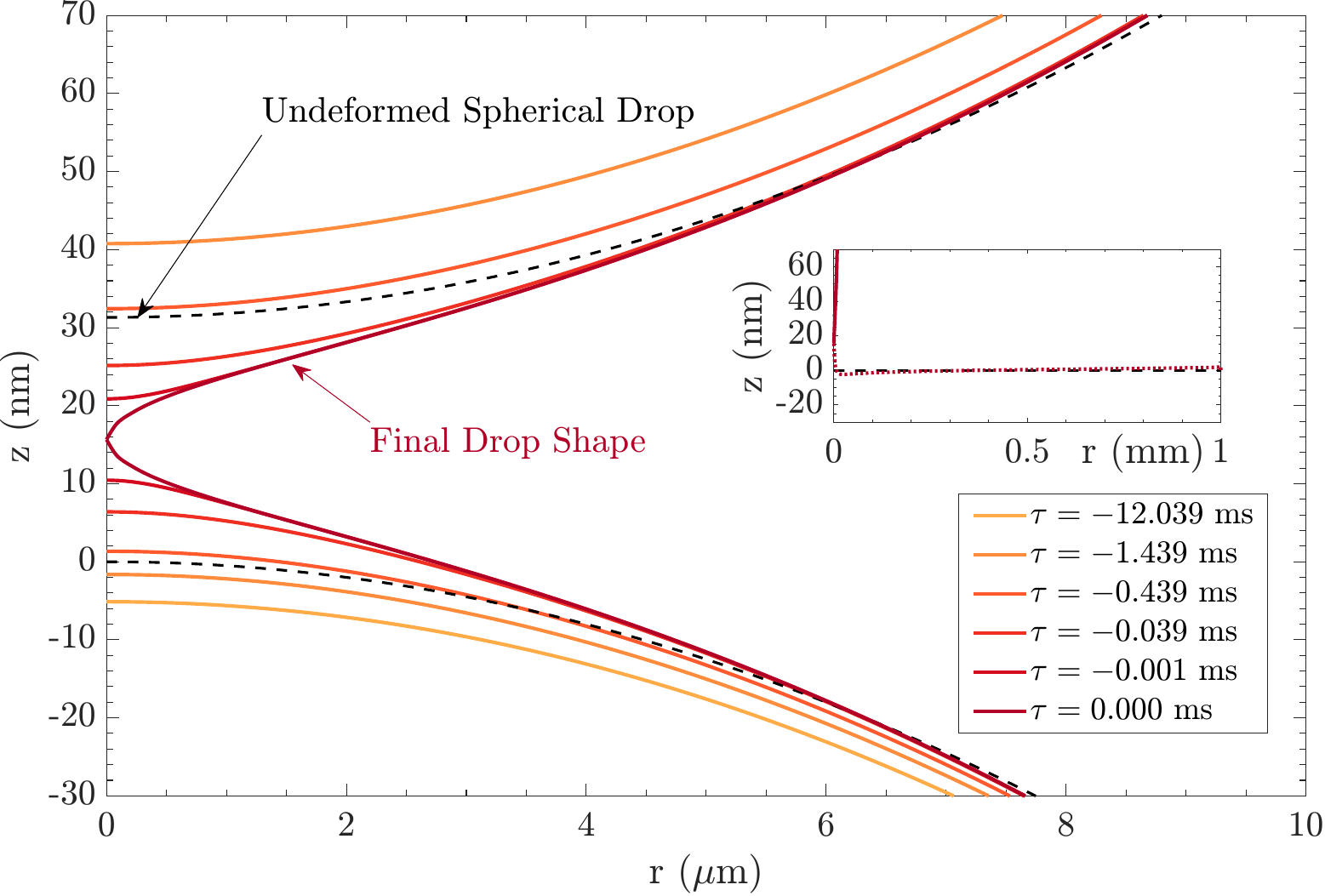}
\caption{Computed interface profiles of the final stages of a drop moving towards another drop at $5~\mu$m/s, just before first contact at $\tau=0$. The inset shows the vertical displacement of the drop from its undeformed spherical shape at $\tau=0$ -- clearly, this is localized near $r=0$. The earliest timestep shown is when the flattening of the lower drop due to the gas film is maximal.}
    \label{fig:DropDrop_fem}
\end{figure}

\begin{figure*}[t]
    \centering
    \includegraphics[width=2\columnwidth]{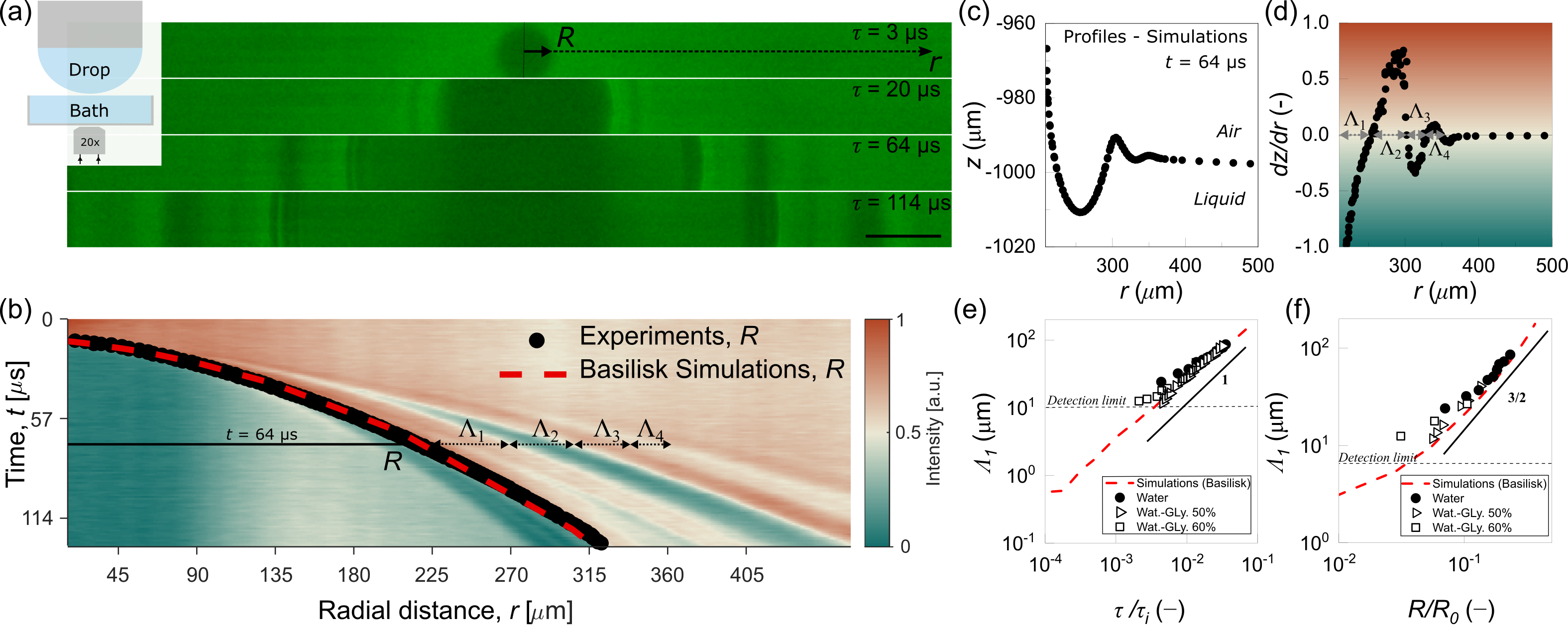}
    \caption{
(a) Image sequences captured from the bottom view using the fast camera. The photographs depict the very early stages of bath interface coalescence, where the droplet (water, $R_0$ = 0.9 mm) contacts the lower bath.
(b) Corresponding spatio-temporal averaged intensity of (a). Black dot points indicate the tracked minimum neck diameter $R$, and the dashed red line represents the Basilisk simulation for the same droplet radius $R_{0}$ as in (a). As the minimum neck diameter increases, capillary waves are excited ahead of the front, as shown by the change in intensity.
(c) Profiles of the interface extracted from the simulation at $\tau$ = 64 $\mu$s after coalescence (left), and its corresponding first derivative (right). We define the lengths $\Lambda_i$ for which the local derivative is positive or negative. These lengths correspond to the change in intensity observed in (b) as a result of the refraction of light at the bath interface.
(e),(f) The first length $\Lambda_1$ corresponding to the air pocket size in (c) as a function of the rescaled time (with the inertial time $\tau_i = \sqrt{\rho R_0^3 / \gamma}$) and the rescaled miminum neck radius.}
    \label{fig:fast_image}
\end{figure*}

Fig.~\ref{fig:electric}(b) shows the minimum neck radius $R/R_0$ as a function of time after coalescence as follows from the electrical measurements: the neck diameter is obtained using relation ~\cite{paulsen2013},
\begin{equation}
R = 2 \left[ 3.62\sigma\left( \Omega - \frac{1}{\pi R_0 \sigma} \right)  \right]^{-1},
\label{eq:radius_electric}
\end{equation}
with $\Omega$ the electrical resistance and $\sigma$ the conductivity of the fluid, which we also confirmed from our simulation (Fig.~S3 in Sup. Mat. \cite{Sup}). The dynamics follow the form $R(\tau) \propto \tau^{\alpha}$, for which we recover $\alpha = 0.5$ for the inertial case. We attribute the different slope observed by Paulsen \textit{et al.}~\cite{paulsen2011} at early times to the initial deformation of the interface due to the presence of charges and the use of a liquid with lower conductivity. To validate this hypothesis, we conducted experiments in which we preliminarily charged the drops and compared the dynamics with drops that remained uncharged. In the presence of charges, side-view images show conical-like deformation (Fig.~S4 in Sup. Mat.) of the interface and cause immediate deviations from the $t^{1/2}$ inertial scaling to $\sim t^{2/3}$ \cite{eddi2013}. However, in the absence of optical imaging at these time scales and spatial scales, we have no information about how the initial contact occurs or the subsequent shape of the interface. These results nonetheless suggest that the details of how the drops are brought into contact can affect the opening dynamics, which will take all its importance below. 

To access this initial pre-contact approach stage, we deploy bespoke finite element simulations for the drop dynamics, which include the influence of the gas, with corrections due to kinetic effects for microscale gas films, alongside van der Waals forces; see Ref.~\cite{sprittles2024} for details of the computational framework. In Fig.~\ref{fig:DropDrop_fem}, we see that for the slow approach speeds considered in our experiments, the deformation of the drops caused by gas lubrication is relatively small so that the interfaces meet at $r=0$, \textit{i.e.} no bubble is entrapped. Notably, the van~der~Waals force causes a `jump to contact' when the separation of the drops is $\approx 30$~nm over a timescale of $\sim$0.1~ms. Separations of tens of namometers are consistent with the literature \cite{ledesma-alonso2012a,ledesma-alonso2012b,chireux2018,chireux2021} and was previously observed for AFM tips that connect to a fluid surface.

\textit{Drop-Bath.} To access the initial stages of coalescence optically, we turn to the drop-bath configuration and use a novel approach by viewing the event from `underneath' to overcome spurious effects due to side imaging \cite{winkels2012,eddi2013}, where it is not possible to measure small contact radii (less than 20 $\mu$m). The setup consists of a drop on a needle to which liquid is supplied slowly by a syringe pump. This drop is facing a bath of the same liquid (Sup.~Fig.~S7, \cite{Sup}). Our needle tip is millimetric and a drop is slowly grown at its tip until it coalesces with the bulk (growing velocity $\leq$ 20 $\mu$m/s, to minimize the effect of air drainage between the droplet and the bulk surface). The viscosity, density and surface tension are given in the Supplementary Table~S1, ~\cite{Sup}. All experiments were performed at room temperature (22$\pm$1~$^{\circ}$C) and atmospheric pressure.

The growth of the neck bridging the droplet with its bulk is recorded from below through the bath using an inverted microscope (Zeiss Axiovert A1) at a frame rate of up to 1,000,000 fps using a high-speed camera (Phantom TMX 7510) allowing us to reach a spatial resolution of typically $0.9\,\upmu$m/pix. The minimum radius of the neck $R$ is followed in time with an image analysis routine (Fig.~\ref{fig:fast_image}). 

\begin{figure}[b]
    \centering
    \includegraphics[width=0.9\columnwidth]{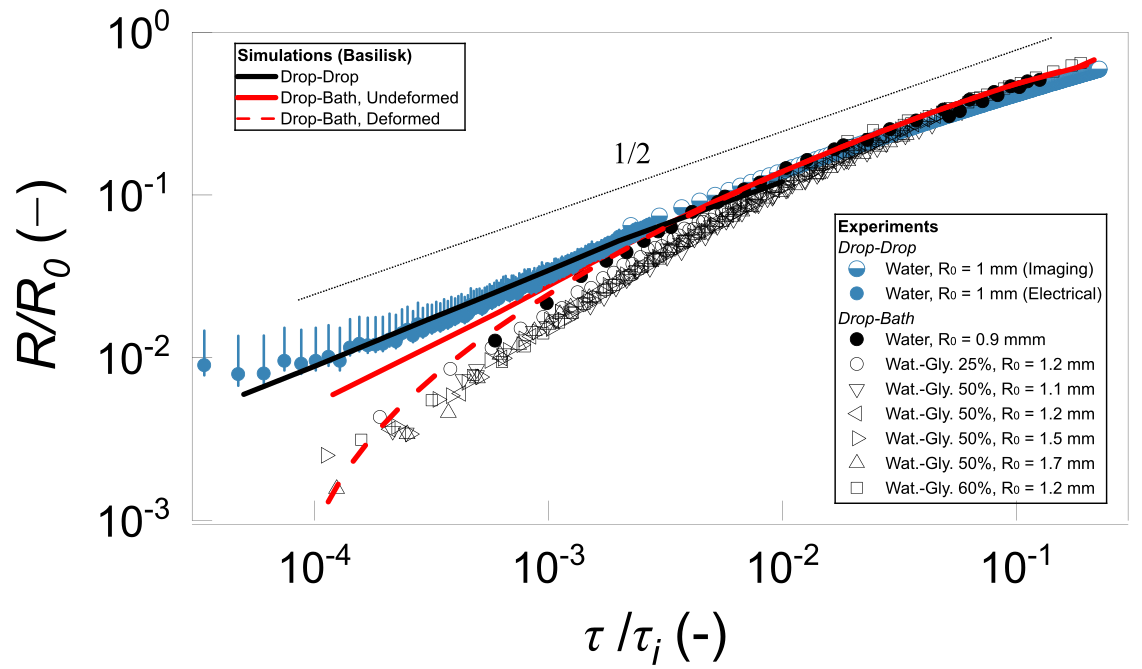}
    \caption{Rescaled radius (with $R_0 = [0.9; 1.73]$ mm) versus rescaled time for the different liquids investigated in the drop-bath configurations, and plotted together with the drop-drop configuration (Fig.~\ref{fig:electric}) for direct comparison. Lines show the results obtained from simulations (Basilisk) and symbols from experiments (imaging and electrical).}
    \label{fig:Dynamics_DropBath}
\end{figure}

Fig.~\ref{fig:fast_image}(a) shows a sequence of the neck opening during coalescence for the droplet-bulk geometry, exhibiting two key features. It is immediately clear that (i) there is an abrupt change of intensity corresponding to the coalesced bridge radius, and (ii) there are concentric rings ahead of the bridge, as seen in Fig.~\ref{fig:fast_image}(b), which appear to be the signature of interfacial waves refracting passing light. This is confirmed by examining the interface profile obtained from the Basilisk simulations at a specific time ($\tau$ = 64 $\mu$s), shown in Fig.~\ref{fig:fast_image}(c), where waves are observed ahead of an air pocket. This combination of inertial coalescence driving wave formation and gas lubrication preventing the pinching of toroidal bubbles has been previously described theoretically and in simulations \cite{duchemin2003, sprittles2014}, and is also predicted for viscous coalescence \cite{eggers1999}, but direct and strong experimental evidence for it was lacking until now.

To further confirm this hypothesis, in Fig.~\ref{fig:fast_image}(d) we plot the spatial deformation ($dz/dr$) obtained from the simulations to define domains $\Lambda_i$ where the interface undergoes pronounced changes. In Fig.~\ref{fig:fast_image}(b), we show that this aligns with experimentally observed intensity variations. The characteristic size of the first domain $\Lambda_1$, which corresponds to the air pocket size shown in Fig.~\ref{fig:fast_image}(c), is shown in [Figs.~\ref{fig:fast_image}(e),(f)] as a function of time and neck radius, respectively. Simulations and experimental results again show good agreement and demonstrate that the air pocket grows linearly in time [Fig.~\ref{fig:Dynamics_DropBath}(e)] and as $R^{3/2}$, as predicted by theory \cite{eggers1999}, regardless of the different viscosities of fluids (1~mPa$\cdot$s to 11.5~mPa$\cdot$s) that we tested in our experiments.

As shown in Fig.~\ref{fig:Dynamics_DropBath}, for the late stage of coalescence, the inertial-capillary regime of drop-drop coalescence is recovered with $R(\tau) \sim \tau^{1/2}$. In the early stage, a different regime is observed for all fluids, with a radius smaller than what would be expected from the inertial-capillary regime: Something else slows down the coalescence. For this earlier regime, all the curves collapse when scaled with the inertial time: There is no effect of viscosity. We therefore look into the dynamics of the air film.

\begin{figure}[t]
    \centering
    \includegraphics[width=\columnwidth]{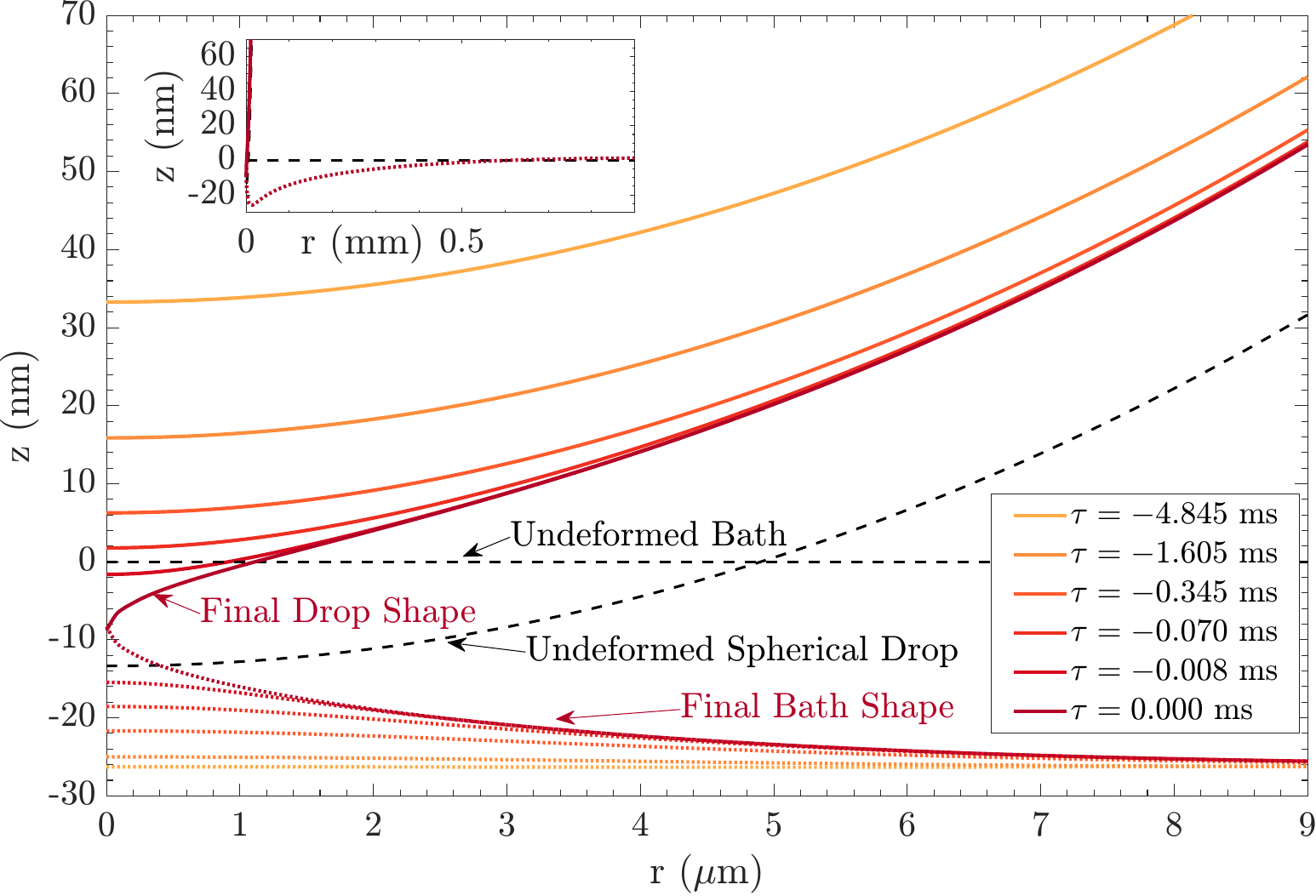}
    \caption{Same graph as in Figure~\ref{fig:DropDrop_fem}, but here we computed the interface profiles of the final stages of a drop moving towards a liquid bath before first contact at $\tau=0$ (approach speed is $5~\mu$m/s). The inset shows the extent of the deformation of the drop and bath at the moment of contact, highlighting differences with the drop-drop case. The earliest timestep shown is when the deformation of the bath due to the gas film is maximal.}
    \label{fig:DropBath_fem}
\end{figure}

In our experiments, we observe interference fringes between the two surfaces in the last frames before coalescence (see Sup.~Fig.~S8), which suggest the presence of a thin air film which acts as a lubricant between the two approaching surfaces. This air film continually thins, and if it becomes thin enough, the two surfaces jump into contact. This behaviour is also confirmed by finite element simulations in the Fig.~\ref{fig:DropBath_fem}, showing the final moments of the approach of a drop towards a liquid bath before coalescence. The gas is modeled using a lubrication equation, and incorporates the van der Waals-driven disjoining pressure that causes the `jump to contact', as well as gas kinetic factors that account for microscale gas films, as described in \cite{sprittles2024}. Fig.~\ref{fig:DropBath_fem} shows that the bath is pushed down vertically $\sim$ 30 nm over the radial extend of the drop $\sim$mm before the drop `jumps to contact', over a radial extent $\sim 3\ \mu$m in $\sim 0.1$ ms, similar to the mechanism seen in the drop-drop case but with a vertical shift.

All this suggest that the initial deformation of the interfaces due to the presence of the air film can change the initial opening dynamics of the liquid bridge, and this effect is more important for the drop-bath than for the drop-drop case, because of the different geometries. To see whether we can reproduce the dynamics of the radius, Basilisk simulations (dashed red line) of the droplet-bulk coalescence for water succeed in qualitatively describing the deviation from the $t^{1/2}$ law; we use an initial neck diameter of 2.7 $\mu$m, which is the smallest size feasible in these simulations. To achieve this agreement, the time axis in the simulation was set to match the experiments [Fig.~\ref{fig:fast_image}(b)]; adjusting the time is rather standard and amounts to fixing the coalescence time $t_0$. If we use the final profiles from the finite-element calculation as starting point, we get a much better agreement (dashed red line, showing again that the initial deformation is important. However, to obtain this agreement, in both the deformed and undeformed cases, we had to adjust the initial minimum neck diameter $R$ by 50 $\mu$m to align the experiments with the simulations (see the comparison in Sup.~Fig.~S9 in Sup.~Mat.). It is perhaps unsurprising that a small spatial shift is required to account for the missed dynamics; in fact, it is rather remarkable that this shift alone is sufficient to achieve good agreement with experiments at later stages. The physical reason behind this shift presents an interesting open question for future work.

Our findings open several exciting avenues for future research. The ultra-fast imaging techniques we deployed, combined with high-precision interferometry, could provide deeper insights into the initial contact dynamics and the evolution of the air film. Additionally, finite element simulations that bridge the pre- and post-contact regimes would offer a more comprehensive understanding of the complex interplay between inertial, capillary, and viscous forces during coalescence. Furthermore, our work has potential applications in microfluidics, inkjet printing, and emulsion stabilization, where precise control over droplet coalescence is crucial. Exploring these applications could lead to innovative advancements in the control of droplet behavior in flows and ensuring film stability across various technologies.

\textit{Acknowledgments.} We are very grateful to the technology center of the university of Amsterdam for technical assistance wit the electrical setup. DB thanks Jacco Snoeijer for very helpful discussions about the artifacts induced by side imaging. JES acknowledges the support of the EPSRC under grants EP/W031426/1, EP/S022848/1, EP/S029966/1, EP/P031684/1 and EP/V012002/1. Peter Lewin-Jones is supported by a studentship within the UK EPSRC–supported Centre for Doctoral Training in the Modeling of Heterogeneous Systems (HetSys), EP/S022848/1. For the purpose of open access, the author has applied a CC BY public copyright license to any Author Accepted Manuscript version arising from this submission.

%

\clearpage
\newpage
\section*{SUPPLEMENTARY MATERIALS}

\section{Drop-drop}

\subsection{Basilisk Simulations.}
Work has been carried out on dimensionless variables using the drop radius, surface tension and density of the liquid as the basis for non-dimensionalizing. In the numerical study, gravitational forces were neglected.
Axisymmetric simulations are performed in a cylindrical coordinate system ($\bar{r}$, $\bar{z}$) in a box of length $L$ where the axis of symmetry lies along the z-axis (the left boundary of the box). 

\begin{figure}[H]
    \centering
    \includegraphics[width=\columnwidth]{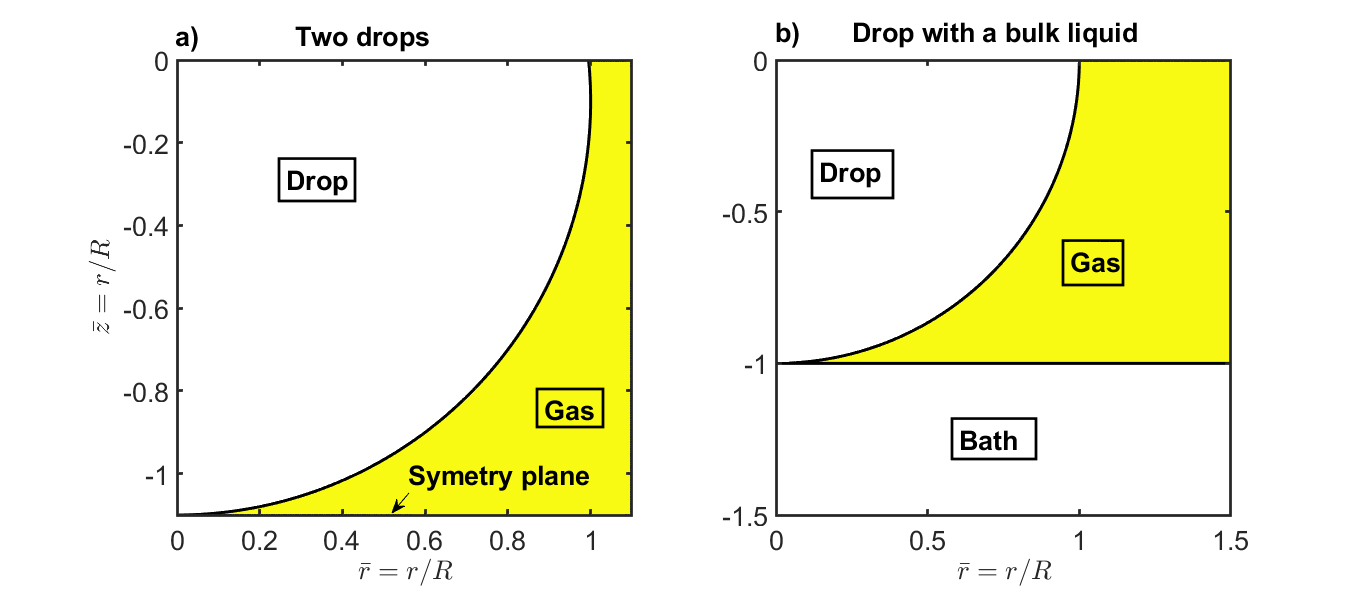}
    \caption{\textbf{Numerical domains}. a) shows the numerical domain for the two drops case while b) It shows the numerical domain used for the drop-bath case.}
    \label{fig:Domains}
\end{figure}

In the 2 drops case, the symmetry of the problem has been exploited to reduce the computational cost. A numerical domain of length L=1.1 was chosen in which one of the drops centered at the point $(0,-0.1)$ was placed [see Fig.~\ref{fig:Domains} (a)]. A symmetry condition was imposed on the bottom boundary to model the other drop. The two drops were brought into contact by an artificial neck with a radius of $0.0028$. Initially, both fluids (liquid and gas) are at rest. The cells in Basilisk are square, so an initial mesh with a maximum refinement of 18 was used in the contact zone to model the collapse of the droplets in the early stages. As the meniscus formed and retracted, this maximum refinement level was reduced.

\subsubsection*{Numerical computation of the electrical resistance}
The numerical resistance was computed using a combination of Basilisk simulations and a finite element method. The methodology used is briefly described below.

First, at the selected time at which the resistance is computed, the numerical profile of the interface is extracted from the Basilisk simulation and saved into a text file. This file is then imported to MATLAB to generate the closed domain needed to solve the Laplace equation for electrical problem. The geometry is normalized with $R_{min}$, and centered at the axial position where the interface has its minimum radius. Neumann boundary conditions are applied on the free surface and on the axis while Dirichlet boundary conditions are applied on the solid supports. A normalized drop in the electrical potential equal to one was considered. Then, the Laplace equation is discretized and solved using a finite element method provided by MATLAB PDE tool. Once we got the electrical potential, the constant and normalized current flow (intensity) through the system is obtained by integrating the axial derivative of electrical potential at any axial position, assuming an electrical conductivity equal to one. Finally, the numerical electrical resistance is simply computed by dividing the intensity by $R_{min}$. See \cite{deblais2018} for further details of the procedure.

\subsection{Finite element simulations}
Whilst Basilisk is able to accurately capture the post-merging coalescence event, volume of fluid techniques are not well suited to resolving the pre-contact `collision' phase to determine the initial contact dynamics.  Here, we use the finite element approach described in detail in \cite{sprittles2024}, focusing it on low-speed collisions as opposed to its usual application to higher speed impacts.  

The computational model is based on solving the Navier-Stokes equations in the bulk of the drop/bath coupled to a lubrication description of the gas phase.  Due to the small film heights encountered, gas kinetic effects are accounted for, and are often crucial, alongside van der Waals forces between the two distinct volumes (drop-drop or drop-bath) that will act to pull the interfaces together. The governing equations are then solved using an open-source finite element package oomph-lib \citep{heil2022} based on the arbitrary Lagrangian Eulerian approach which, crucially, permits a high accuracy representation of the free surface and is well suited to multiscale problems like this one.  Complete details are provided in \cite{sprittles2024}. 

Here, we take the experimental values, using material parameters for the water-air system alongside a collision speed of $5\times10^{-6}~$m/s and, due to its negligible effect on the `jump to contact', neglect gravity. In Fig.~2 the profiles of the interface shape as the two drops approach show that there is very little deformation due to the gas but that the van der Waals forces between the two interfaces causes a `jump to contact' when their separation is $\approx 30$ nm which deformed the drop on a lateral scale in the order of of $\mu$m.

\subsection{Electrical setup}

We measured the resistance of the opening of the neck neck with an electrical circuit analogous to the one described in the references~\cite{Case2009,paulsen2011,paulsen2012}.

\begin{figure}[H]
    \centering
    \includegraphics[width=1\columnwidth]{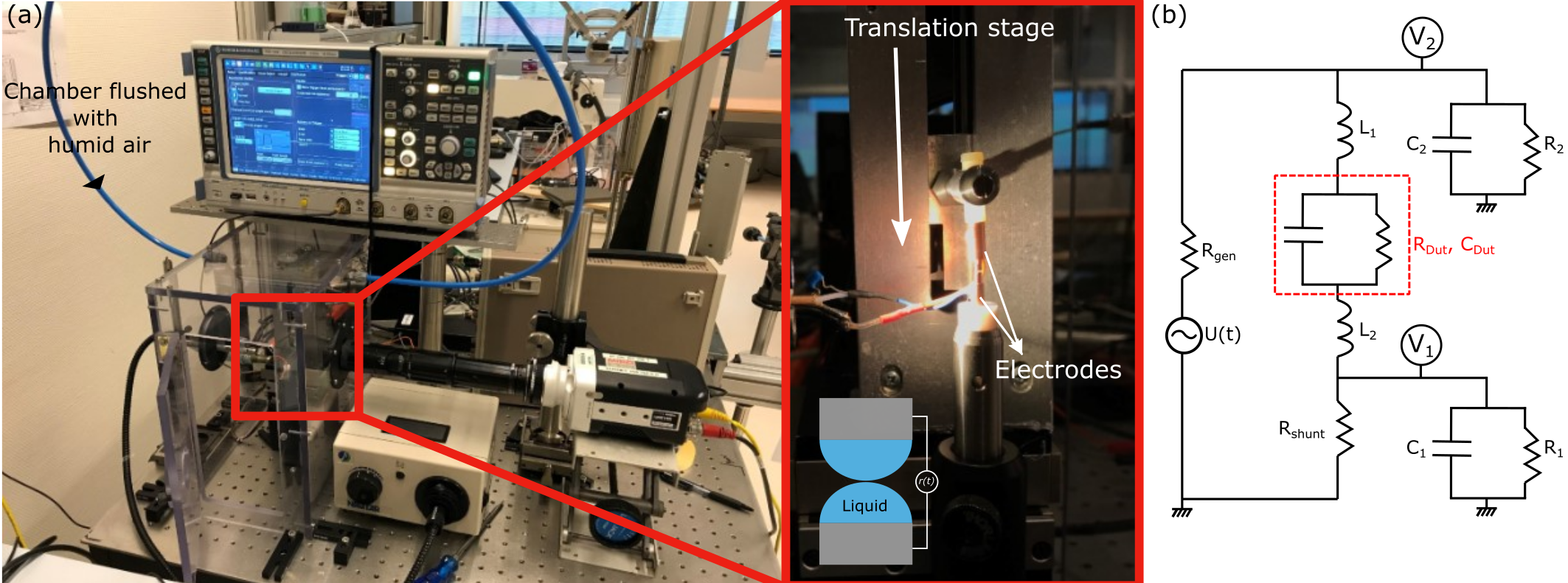}
    \caption{\textbf{Electrical setup used for the measure of the electrical resistance of the coalescence event.} (a) The entire setup is placed in a chamber flushed continuously with humid air to prevent evaporation. A close view of the two electrodes receiving the two droplets. The upper electrode can be translated downward thanks to a speed-controlled translation stage. The dynamics of coalescence is followed, over a large time scale, by measuring the electrical resistance (oscilloscope on the top) and by imaging (fast camera) the capillary bridge. (b) Equivalent electrical circuit.}
\end{figure}

The voltages $V_1$ and $V_2$ were measured and sampled simultaneously at a rate of 1~GHz. The frequency $f$ of the input sine wave from the function generator was set at a maximum value of $f = 10$~MHz, allowing us to reach sub-microsecond time scales. At these frequencies, electromagnetic waves can interfere with the measurements, so we shielded and enclosed the experiment to minimize parasitic capacitances.

The high-frequency voltages were read into Matlab and analyzed using a custom-built routine. The analysis averaged the incoming signals over a single period to determine the ratio of their amplitudes, $V_2/V_1$, as a function of time $t$. In addition, the analysis compared the input signals with a reference sine wave to calculate the relative phase shift of each signal over time. We limited the maximum applied voltage to 1~V to prevent deformation of the droplet interfaces during contact. This 1~V limit has been previously reported in \cite{paulsen2013}, ensuring that the interfaces remain undeformed by the electric field when in close proximity. To avoid such deformation, the measurements reported in Fig.~1 were obtained with an applied voltage of 300~mV.

Furthermore, we observed that external factors, such as the presence of charges, strongly affect the initial shape of the droplets and thus the coalescence dynamics. To minimize this effect, we grounded the setup and used an antistatic gun (Zerostat) before each experiment. Sup.~Fig.~\ref{SupFig:EffectOfCharges} demonstrates how the presence of initial charges influences the dynamics (see the conical shape resulting from the presence of charges in Fig.~\ref{SupFig:EffectOfCharges}).

\subsubsection{Relation between the electrical resistance and the neck diameter}

\begin{figure}[H]
    \centering
    \includegraphics[width=\columnwidth]{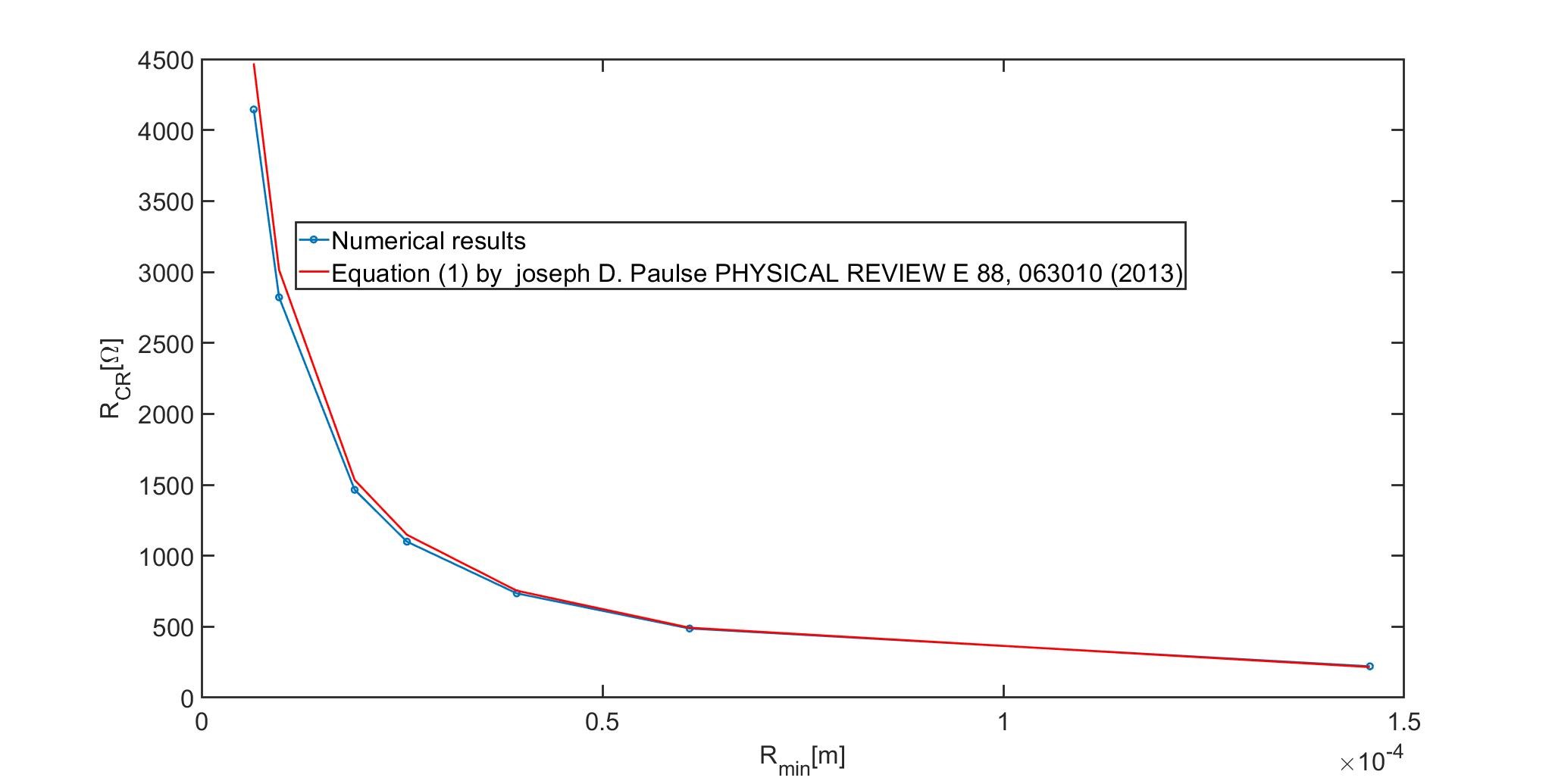}
    \caption{\textbf{Relation between electrical resistance and the minimum neck diameter}. The red solid line are results obtained from the Eq.~\ref{eq:Relation_ElecResistance_NeckDiameter} and the blue solid line is obtained from our Basilisk simulations.}
    \label{fig:Resistance_SimVsEquationPaulsen}
\end{figure}

The minimum neck diameter $R$ is obtained indirectly by measuring the electrical resistance $\Omega$ of the neck in time. To obtain the minimum neck diameter, we use the equation~1 proposed by \cite{paulsen2013}:
\begin{equation}
R=2{[(\Omega-\frac{1}{\pi R_0 \sigma}) 3.62\sigma]}^{-1}
\label{eq:Relation_ElecResistance_NeckDiameter}
\end{equation}
We confirmed the validity of this equation with our simulations by plotting the electrical resistance obtained from our Basilisk simulation as a function of the minimum neck diameter. The results presented in Fig.~\ref{fig:Resistance_SimVsEquationPaulsen} confirmed the good agreement between the two. Thus, we use this equation to convert the measured electrical resistance from our experiments to the minimum physical diameter of the neck $R$.

\subsubsection{Effect of charges and high voltage on the drops' interface and coalescence dynamics.}

\begin{figure}[H]
    \centering
    \includegraphics[width=0.9\columnwidth]{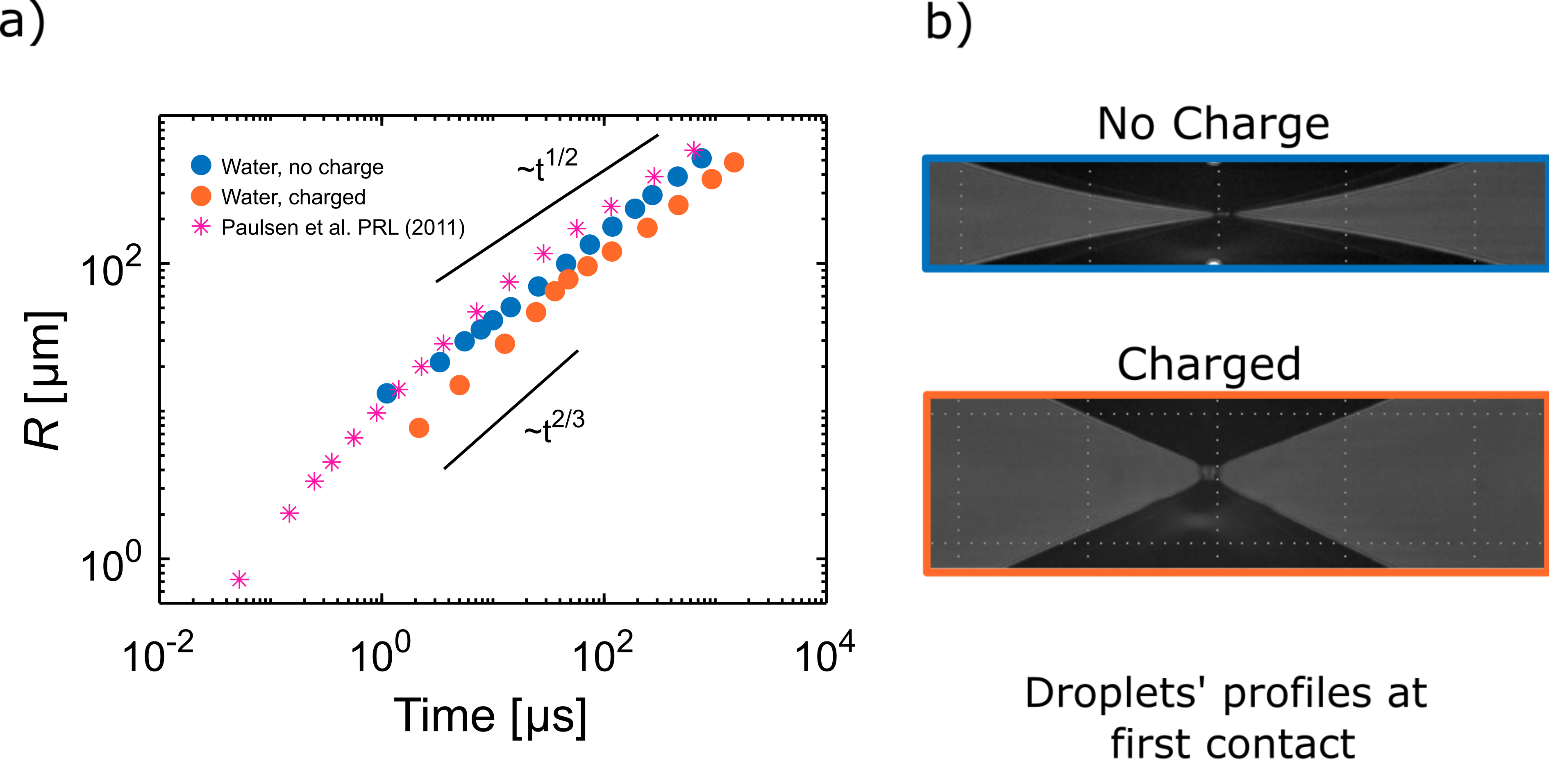}
    \caption{\textbf{Effect of surface charges on the dynamics of coalescence of water drops (imaging).} (a) The presence of charges on water droplets (orange symbols) is shown to influence the coalescence dynamics at early stages, compared to the uncharged case (blue symbols). Note that, the initial drop sizes in these two experiments are slightly different.  (b) Snapshots of the profiles at the first contact of coalescence, highlighting the difference between the two cases.}
    \label{SupFig:EffectOfCharges}
\end{figure}

Fig.~\ref{SupFig:EffectOfCharges} compares charged and uncharged water drops as they approach each other. Initially, two water drops with similar sizes formed on the tips of two vertical plastic nozzles (diameter 2.3 mm). Similar to the setup used in the electric measurement, we fixed the bottom drop while moving the top drop towards the bottom one. Fast imaging was applied from the side using an optical zoom lens (20$\times$ magnification). To charge the drops, we simply used a charged plastic tube (by rubbing with dry hair) to get close to the drops before coalescence occurred.

When the drops are charged, a conical liquid neck appears in the early moments of coalescence (Fig.~\ref{SupFig:EffectOfCharges}b). This charge effect causes a deviation in dynamics from the noncharged case, following the capillary-inertial scaling, $R \sim t^{1/2}$ to $R \sim t^{2/3}$ (Fig. \ref{SupFig:EffectOfCharges}a). We believe that this pointed shape of drops could also occur in drop-drop electric experiment where an external electric field is applied, such as in the work of \cite{paulsen2011}. However, in our drop-drop electric measurement, before triggering coalescence, we carefully used a zero-static gun to discharge the drops, ruling out the pointed shapes.

\subsubsection{Comparison with Beaty and Lister (2022)}

\begin{figure}[h]
    \centering
    \includegraphics[width=0.7\columnwidth]{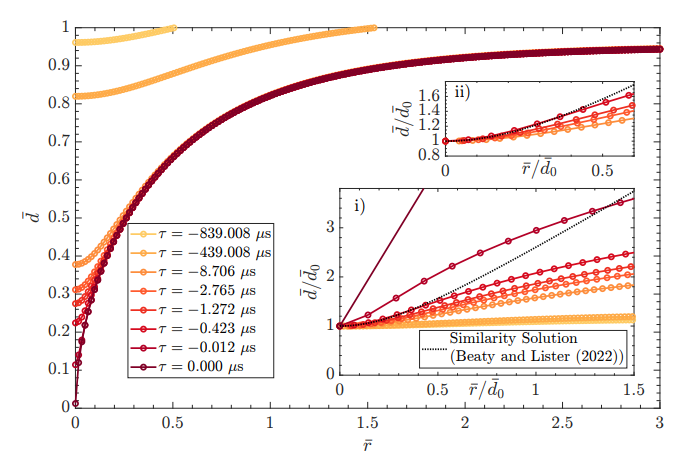}
    \caption{A comparison between our drop-drop simulations in Figure~3, and the similarity solution predicted by Beaty and Lister (2022).}
    \label{SupFig:BeatyLister}
\end{figure}

Beaty and Lister (2022) developed an analytical theory for the van der Waals-driven contact of two drops. This neglects inertia, and the initial condition is of stationary drops perturbed from a spherical state, rather than spherical drops moving slowly, as in our simulations. They also neglect the effects of the surrounding gas. They construct a similarity solution as follows. They define the non-dimensional parameter $\mathcal{H}=\frac{A}{3\pi \gamma R^2}$, which characterizes the relative strength of the van der Waals force compared to the surface tension. The rescaled, non-dimensional radial coordinate is defined as $\bar{r}=r/(R\mathcal{H}^{1/6})$, and the perturbed gap width as $\bar{d}(r,t)=(2z(r,t)-r^2/R)/(R\mathcal{H}^{1/3})$. This rescaling is shown in the main plot of Fig.~\ref{SupFig:BeatyLister}, where we note qualitative agreement with Fig~2 of Beaty and Lister (2022). The circles represent the mesh points of our simulation. In inset (i), we rescale by $\bar{d}_0=\bar{d}(0,t)$ and compare to the similarity solution $(\bar{d}_0+\theta^2\bar{r}^2)^{1/2}$, where $\theta$ depends on the initial condition, given by $\frac{\partial{\bar{d}}}{\partial{\bar{r}}}$ at the edge of the validity of the similarity solution. For this comparison, we fit $\theta$. The plot shows that we approach the similarity solution (shown zoomed in inset (ii)), but in the final moments of approach, the resolution of our mesh becomes too coarse (due to the division by $d_0$), and the simulations no longer converge to the similarity solution.
The results in Beaty and Lister (2022) neglect inertia. In their second paper (Beaty and Lister, 2023), they include inertia, but we have been unable to achieve agreement with their results for the viscous-inertial regime.

\section{DROP-BULK}

\subsection{Basilisk Simulations} 
In this case, a computational domain of length L = 1.5 has been chosen in which the droplet is centered at the point (0,0). A liquid bath with a flat and free surface was added at a height $\hat{z}=z/R=-1$ (see Fig. S1(b)). The drop and the bath were brought into contact by an artificial neck with a radius of $0.003$. An initial mesh with a maximum refinement of 19 was used in the contact zone to model the collapse of the droplet with the bath in the early stage. Again, as the meniscus formed and retracted, this maximum refinement level was reduced. 

\begin{figure}[H]
    \centering
    \includegraphics[width=0.9\columnwidth]{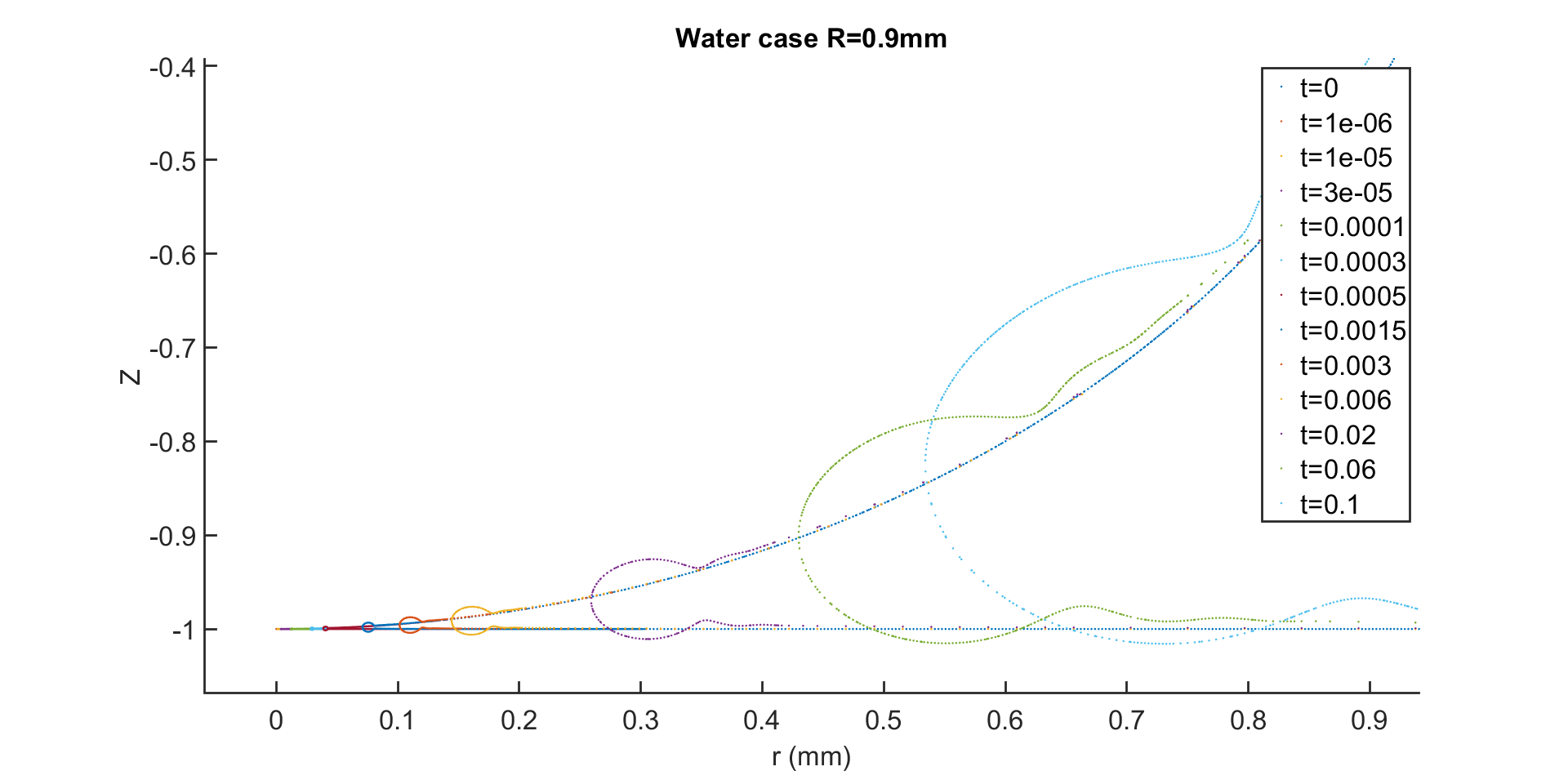}
    \caption{Simulated (Basilisk) interface profiles for different times after coalescence in the drop-bulk configuration. Note the formation and the growth of the air pocket followed by waves.}
    \label{fig:exp_setup}
\end{figure}

An additional simulation (labeled `Deformed' in Fig.~4 of the main text) instead used the bath profile computed at the moment of contact by the finite element simulation in Fig.~5, where the bath has been forced down $\sim 30$ nm. The spherical drop was also shifted vertically down so that at $r=0$ the vertical distance between the drop and the bath is the same as that found by the finite element simulation before the `jump to contact' occurred ($t=-0.07$ ms in Fig.~5) An artificial neck of radius 0.003 was then added (the same as the `undeformed' case).

\subsection{Finite element simulations} 
Fig.~5 for the drop-bath case in the main text shows that the drop exhibits a rather similar shape to those recovered for the drop-drop case, with a notable `jump to contact'. However, the deformation of the bath is more prominent, with a `dip' of $\sim 30$ nm below its initial height seen when contact occurs.  This is because the Laplace pressure associated with the planar, or slightly deformed, bath interface is significantly lower than the drop's, and therefore the gas lubrication pressure has more of an influence on it -- one can see from the figure that this deformation has occurred well before the final stages of contact, and the inset shows that it extends radially to the width of the drop.
\bigskip

\begin{figure}[H]
    \centering
    \includegraphics[width=0.7\columnwidth]{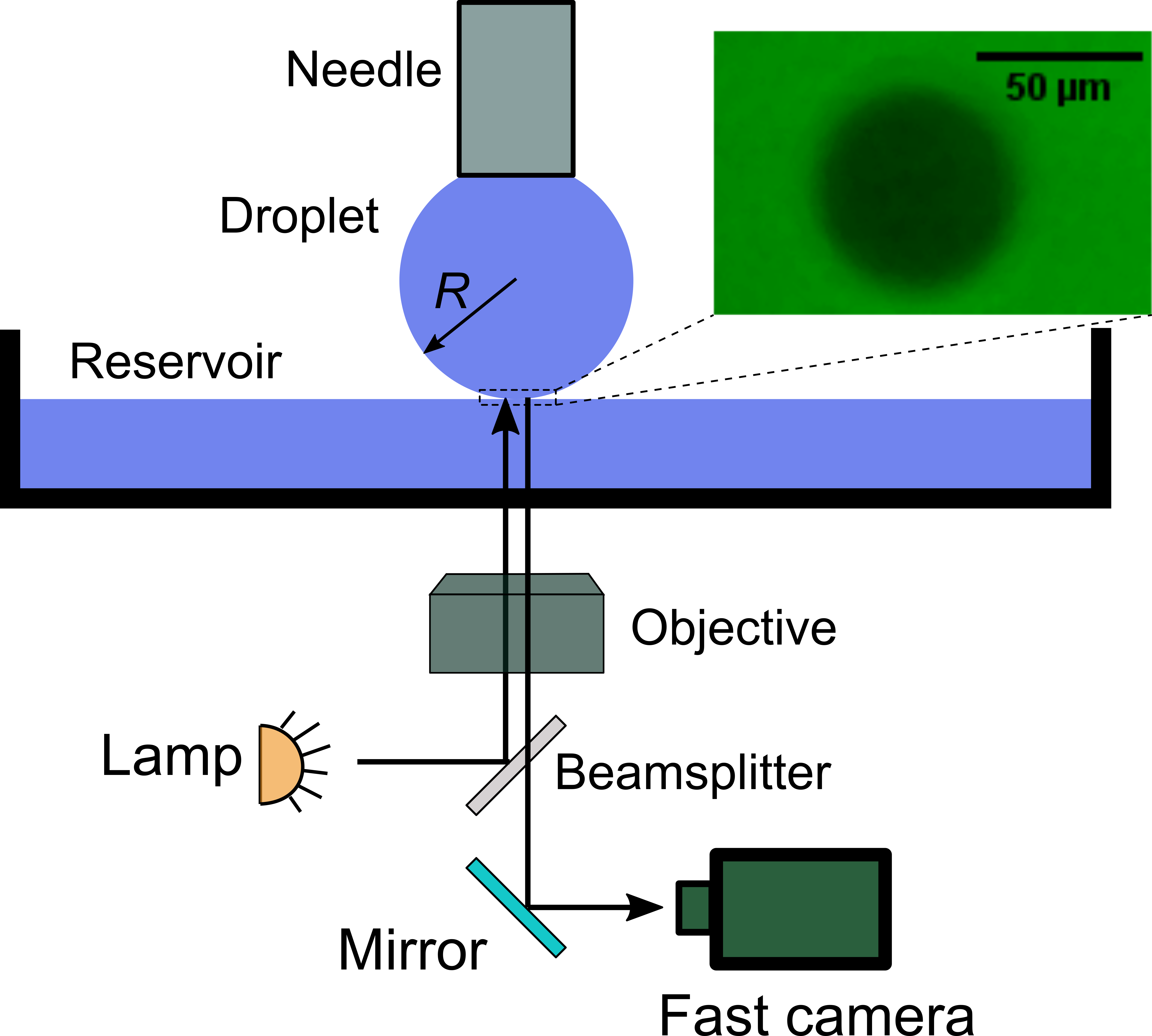}
    \caption{Experimental setup for drop-bulk coalescence. The setup combined a microscope (Zeiss Observer) coupled to a fast camera with different magnifications.}
    \label{fig:exp_setup}
\end{figure}

\subsection{Imaging setup}
As mentioned in the main text, to access the initial stages of coalescence optically, we deploy a novel approach by viewing the event from `underneath' to overcome spurious effects due to the side imaging (\cite{winkels2012,eddi2013}). The setup consists of a drop on a needle to which liquid is supplied slowly by a syringe pump (Harvard Apparatus). This drop is facing a bath of the same liquid (See Fig.~\ref{fig:exp_setup}). Our needle tip is typically millimetric (diameter 1.37 mm) and a drop is slowly grown at its tip until it coalesces with the bulk (growing velocity $\leq$ 20 $\mu$m/s, in an attempt to minimize the effect of air drainage between the droplet and the bulk surface). To change the properties of the fluids, we mixed different amounts of glycerol with deionized water (Milli-Q). The viscosity, density and surface tension are given in the Supplementary Table~S1. All experiments were carried out at room temperature (22$\pm$1~$^{\circ}$C) and atmospheric pressure.

The growth of the neck bridging the droplet with its bulk is recorded from below through the bath using an inverted microscope (Zeiss Axiovert A1) at a frame rate of up to 1,000,000 frames per second (fps) using a high-speed camera (Phantom TMX 7510) allowing us to reach a spatial resolution of typically 0.9 $\mu$m/pix with a 20$\times$ microscope objective (Zeiss long working distance WD = 20 mm).

\begin{figure}[H]
    \centering
    \includegraphics[width=0.8\columnwidth]{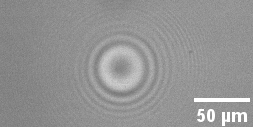}
    \caption{\textbf{The fringes before the drop coalescing into the bulk.} }
    \label{SupFig:InterferenceFringes}
\end{figure}

Before coalescence, we observed interference fringes between the two surfaces in the last instances (see Fig.~\ref{SupFig:InterferenceFringes}), which suggest the presence of a thin air film that acts as a lubricant between the two approaching surfaces. This air film continually thins, and if it becomes thin enough, the two surfaces jump into contact. 

\begin{figure}[H]
    \centering
    \includegraphics[width=1\columnwidth]{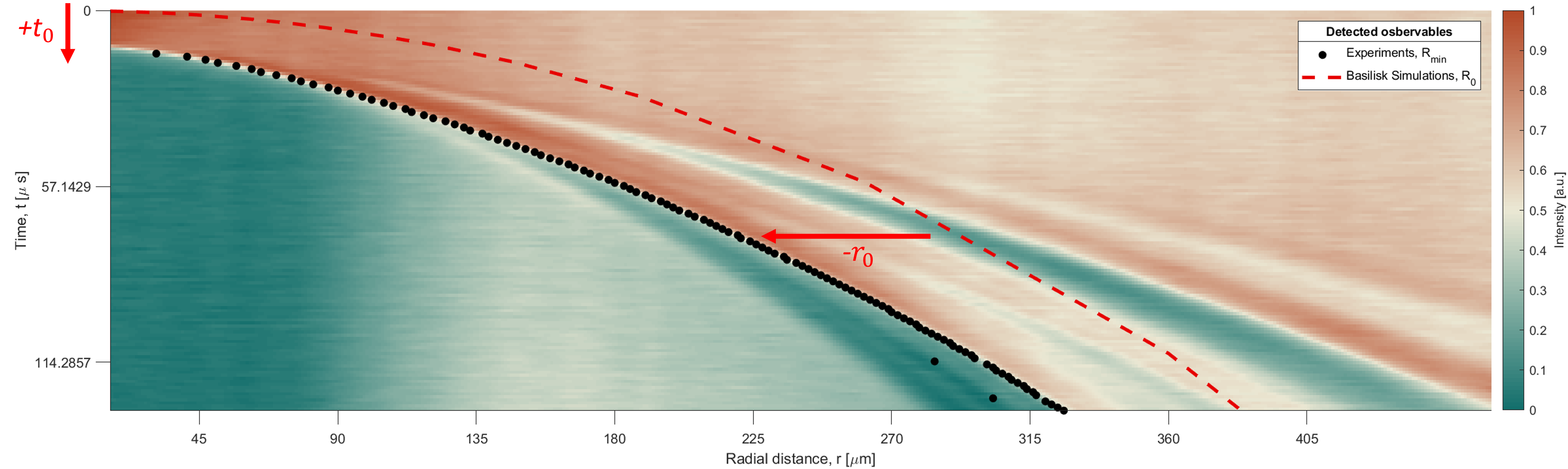}
    \caption{Saptio-temporal averaged intensity compared to the Basilisk simulation without the artificial shift ($t_0, r_0$).}
    \label{SpatioTemporal_Graph_Shift}
\end{figure}

Table \ref{tab:fluids} shows the properties of the fluids used in this study. Different glycerol fractions (by mass) were mixed with deionized water to tune the viscosity. 

\begin{center}
Liquid properties used in the electrical (drop-drop) and imaging (drop-bulk) experiments.
\begin{tabular}{c|c|c|c|c|c}
   \hline
  \begin{tabular}{c} Fluids \end{tabular} & 
  \begin{tabular}{c} Glycerol fraction \\ $\phi$ [wt\%] \end{tabular} & 
  \begin{tabular}{c} Viscosity \\ $\eta$ [mPa$\cdot$s] \end{tabular} &
  \begin{tabular}{c} Surf. Tension \\ $\gamma$ [mN/m] \end{tabular} & 
  \begin{tabular}{c} Density \\ $\rho$ [g/mL] \end{tabular} &
  \begin{tabular}{c} Conductivity \\ $\sigma$ [mS/cm] \end{tabular} \\
  \hline
    Water + NaCl & 0 & 1.78 & 82 & 1.12 & 30\\
   Water &  0  &  1 & 72 & 1 & - \\
   Water+Glycerol &  25  &  2.16 & 71 & 1.06 & - \\
   Water+Glycerol &  50  &  5.71 & 69 & 1.13 & -\\ 
   Water+Glycerol &  60  &  11.5 & 68 & 1.15 & -\\
   \hline
\end{tabular}
\label{tab:fluids}
\end{center}

%

\end{document}